\PassOptionsToPackage{table}{xcolor}
\documentclass[conference]{IEEEtran}

\usepackage{hyperref}
\usepackage{cite}
\usepackage{amsmath,amssymb,amsfonts,pifont}
\usepackage{textcomp}
\usepackage[most]{tcolorbox}
\usepackage[inline]{enumitem}

\usepackage{graphicx}
\usepackage{booktabs}
\usepackage{multirow}
\usepackage{array}
\usepackage{balance}
\usepackage{makecell}
\newcolumntype{T}{>{\raggedleft\arraybackslash}p{1.5cm}} 
\newcolumntype{V}{>{\centering\arraybackslash}p{0.2cm}}  

\usepackage[table]{xcolor}
\definecolor{brilliantrose}{rgb}{1.0, 0.33, 0.64}
\definecolor{classicrose}{rgb}{0.98, 0.8, 0.91}
\definecolor{gamboge}{rgb}{0.89, 0.61, 0.06}

\definecolor{darkpink}{rgb}{0.91, 0.33, 0.5}

\definecolor{properties-yaml}{rgb}{0.54, 0.17, 0.89}
\definecolor{text-yaml}{rgb}{0.0, 0.0, 0.0}
\definecolor{comment-yaml}{rgb}{0.36, 0.22, 0.33}

\usepackage{comment}
\usepackage{xspace}
\usepackage{ifthen}
\newboolean{showcomments}
\setboolean{showcomments}{false} 
\ifthenelse{\boolean{showcomments}}{ 
  \newcommand{\nbnote}[3]{
	\fcolorbox{brilliantrose}{classicrose}{\bfseries\sffamily\scriptsize#1}{
	  \color{#2} 
	  \sffamily
	  \small
	  $\blacktriangleright$
	  \textit{#3}
	  $\blacktriangleleft$
	}
  }
}{
  \newcommand{\nbnote}[3]{}
  
}

\newcommand{\code}[1]{%
  {\ttfamily\mdseries\selectfont\codeprocess#1\relax
  }%
}

\def\codeprocess#1{%
  \ifx#1\relax
  \else
    \ifx#1\_%
      \scalebox{0.5}[1]{\textunderscore}
    \else
      #1%
    \fi
    \expandafter\codeprocess
  \fi
}

\usepackage{xurl}
\usepackage{listings}
\usepackage{algorithm, algorithmicx}
\usepackage[noend]{algpseudocode}
\usepackage{etoolbox}
\algrenewcommand\algorithmicindent{0.7em} 
\AtBeginEnvironment{algorithm}{\small}    
\AtBeginEnvironment{algorithmic}{\small}  

\algnewcommand\algorithmicswitch{\textbf{switch}}
\algnewcommand\algorithmiccase{\textbf{case}}

\algdef{SE}[SWITCH]{Switch}{EndSwitch}[1]{\algorithmicswitch\ #1}{\algorithmicend\ \algorithmicswitch}
\algdef{SE}[CASE]{Case}{EndCase}[1]{\algorithmiccase\ #1}{\algorithmicend\ \algorithmiccase}
\algtext*{EndSwitch}
\algtext*{EndCase}

\definecolor{purple-call}{rgb}{0.54, 0.17, 0.89}
\algrenewcommand\Call[2]{\textcolor{purple-call}{\textproc{#1}}(#2)} 
\algrenewcommand\textproc[1]{\textcolor{purple-call}{\textsc{#1}}}

\newcommand{\icepick}{{\fontfamily{ptm}\selectfont\textsc{IcePick}}\xspace} 
\newcommand{\glacier}{{\fontfamily{ptm}\selectfont\textsc{Glacier}}\xspace}
\newcommand{\rtg}{{\fontfamily{ptm}\selectfont\textsc{RestTestGen}}\xspace}

\newcommand{\success}{\code{2xx}\xspace}
\newcommand{\clienterr}{\code{4xx}\xspace}
\newcommand{\error}{\code{5xx}\xspace}

\newcommand{\notfound}{\code{404}\xspace}
\newcommand{\ok}{\code{200}\xspace}

\newcommand{\tla}{TLA$^{+}$\xspace}

\makeatletter

\usepackage[T1]{fontenc}
\usepackage{inconsolata}

\newcommand\YAMLkeystyle{%
  \ttfamily\slshape\footnotesize\color{properties-yaml}
}
\newcommand\YAMLvaluestyle{%
  \ttfamily\slshape\footnotesize\color{text-yaml}
}
\newcommand\LSTNumberStyle{%
  \ttfamily\slshape\footnotesize\color{black}
}

\makeatletter
\newcommand\language@yaml{yaml}

\expandafter\expandafter\expandafter\lstdefinelanguage
\expandafter{\language@yaml}
{
  keywords={true,false,null},
  showstringspaces=false,
  backgroundcolor=\color{white},
  stepnumber=1,
  upquote=true,
  keywordstyle=\color{text-yaml},
  basicstyle=\YAMLkeystyle,
  breaklines=true,                                 
  sensitive=false,
  comment=\color{comment-yaml}[l]{\#},
  morecomment=[s]{/*}{*/},
  commentstyle=\color{comment-yaml},
  stringstyle=\YAMLvaluestyle,
  moredelim=[l][\color{orange}]{\&},
  moredelim=[l][\color{magenta}]{*},
  moredelim=[is][\YAMLvaluestyle]{+}{+},
  columns=flexible,
  tabsize=1,
  numbers=left,
  numberstyle=\LSTNumberStyle,
  xleftmargin=2em,        
  framexleftmargin=2em,   
}

\lst@AddToHook{EveryLine}{\ifx\lst@language\language@yaml\YAMLkeystyle\fi}
\makeatother

\definecolor{purple-tla}{rgb}{0.54, 0.17, 0.89}
\definecolor{blue-tla}{rgb}{0.16, 0.32, 0.75}
\definecolor{green-tla}{rgb}{0.13, 0.55, 0.13}

\newcommand\TLAKeyword{%
  \ttfamily\slshape\footnotesize\color{purple-tla}\selectfont
}
\newcommand\TLAText{%
  \ttfamily\slshape\footnotesize\color{black}}%
  \newcommand\TLALineComment{\ttfamily\slshape%
  \footnotesize\color{green-tla}\selectfont
}
\newcommand\TLAMultilineComment{%
  \ttfamily\slshape\footnotesize\color{blue-tla}\selectfont
}
\newcommand\TLAString{%
  \ttfamily\slshape\footnotesize\color{blue-tla}\selectfont
}
\newcommand\TLABoolean{%
  \ttfamily\slshape\footnotesize\color{blue-tla}\selectfont
}

\makeatletter
\newcommand\language@tla{tla}

\expandafter\expandafter\expandafter\lstdefinelanguage
\expandafter{\language@tla}
{
  keywords=[1]{
    VARIABLES, ASSUME, CONSTANTS, SUBSET, IF, 
    THEN, ELSE, CHOOSE, LET, IN, UNCHANGED, EXCEPT, 
    in, notin, DOMAIN, EmptyMap, MapPut, MapDel, IsMap
  },
  keywordstyle=[1]\TLAKeyword,
  keywords=[2]{TRUE, FALSE},
  keywordstyle=[2]\TLABoolean,
  showstringspaces=false,
  numbers=left,  
  backgroundcolor=\color{white},
  numberstyle=\LSTNumberStyle, 
  stepnumber=1,
  upquote=true,
  basicstyle=\TLAText,
  breaklines=true,                                 
  sensitive=false,
  comment=\TLALineComment[l]{\\*},
  morecomment=\TLAMultilineComment[s]{\(*}{*\)},
  morestring=[b]",
  stringstyle=\TLAString,
  xleftmargin=15pt, 
  tabsize=1, 
  basewidth=0.55em,
  literate=
    {/\\}{{$\land$}}1
    {|->}{{$\mathord{\mapsto}$}}1
    {<<}{{$\langle\!\langle$}}1
    {>>}{{$\rangle\!\rangle$}}1
    {\\notin}{{$\mathbin{\not\in}$}}1
    {\\in}{{$\mathbin{\in}$}}1
    {\\A}{{$\forall$}}1
    {\\E}{{$\exists$}}1
    {=>}{{$\Rightarrow$}}1
    {<=}{{$\leq$}}1
    {\{\}}{{$\varnothing$}}1
    {\#}{{$\mathbin{\neq}$}}1
}

\lst@AddToHook{EveryLine}{\ifx\lst@language\language@tla\YAMLText\fi}
\makeatother

\newtcolorbox{callout}{
  enhanced,
  boxrule=0.6pt,      
  arc=2mm,            
  left=0.2mm,right=0.5mm,top=0.5mm,bottom=0.5mm,
  boxsep=0.5mm,
  colback=white,      
  colframe=black,
}

\begin{document}

\title{Systematic API Testing Through Model Checking and Executable Contracts}
\author{\IEEEauthorblockN{Ana Catarina Ribeiro}
\IEEEauthorblockA{NOVA University Lisbon \\
Caparica, Portugal \\
acm.ribeiro@campus.fct.unl.pt}
\and
\IEEEauthorblockN{Margarida Mamede}
\IEEEauthorblockA{NOVA University Lisbon \\
Caparica, Portugal \\
mm@fct.unl.pt}
\and
\IEEEauthorblockN{Carla Ferreira}
\IEEEauthorblockA{NOVA University Lisbon\\
Caparica, Portugal \\
carla.ferreira@fct.unl.pt}
}

\maketitle

\begin{abstract}

Automated black-box testing of APIs typically relies on interface specifications that define available operations and data schemas, but offer limited or no behavioural semantics. This semantic gap amplifies the test-oracle problem and limits the generation of effective, stateful call sequences.

We introduce \icepick, a framework that achieves systematic state-space coverage for API testing by leveraging model checking. \icepick uses \tla to formally model API state evolution, employs the TLC model checker to exhaustively explore reachable states, and generates test sequences that provably cover the behavioural model. To mitigate state-space explosion and improve sequence extraction, we introduce a coverage-guided breadth-first traversal of the TLC state-space graph.
To address oracle limitations beyond HTTP status codes, we propose \glacier, a first-order logic contract language that enriches API specifications with executable semantic contracts, enabling automated behavioural verification during test execution.

We evaluate \icepick on EvoMaster Benchmark systems, demonstrating that model-checking-guided exploration achieves complete state coverage and reveals faults in multi-operation interactions. We also analyse scalability to characterise practical limits and applicability requirements. Overall, \icepick provides reproducible test suites with strong coverage guarantees for critical API-based systems.

\end{abstract}


\section{Introduction}
\label{sec:introduction}
Model checking~\cite{model-checking} and testing are complementary approaches to establishing system correctness. Model checking provides formal guarantees by exhaustively exploring a model's complete state-space to verify specific properties. Testing, by contrast, validates the actual implementation by executing it with diverse inputs to \mbox{uncover edge cases~\cite{pgodefroid,generate-from-spec}}.

In this paper, we integrate both approaches by leveraging model checking's systematic state-space exploration to generate test cases that comprehensively cover a program's behaviour in a black-box setting. We apply this combined approach to microservice architectures, which decompose business logic into a suite of small services~\cite{fowler}. Because such systems are often accessed by heterogeneous third-party clients, their implementations are typically hidden, making them opaque to users. Most microservice implementations follow REST, a set of architectural principles for building web services over HTTP~\cite{evomaster1}, and their APIs are commonly referred to as RESTful APIs. While we focus on microservice architectures, the approach applies to other API-based systems that are testable through their interfaces.

Testing microservices poses unique challenges, as the main artefact is the API definition, typically specified using the OpenAPI Specification (OAS)~\cite{oas}. While OAS is widely accepted as a standard, it lacks detail to fully capture the system behaviour. These specifications outline operation endpoints, expected data types for each HTTP interaction, and response codes for various scenarios, yet they fall short of capturing the service's underlying logic and dynamic behaviours.

The widespread adoption of RESTful APIs has driven demand for automated black-box testing tools.
Existing tools typically derive tests exclusively from API specifications and face several limitations despite promising results~\cite{evomaster2, restestgen, morest, restler, restest, restct, agora, spec-based, metamorphic}. Specifications are often incomplete (missing parameters or dependencies), inconsistent (mismatched names across operations), outdated (not synchronised with implementation), or non-compliant with RESTful guidelines. However, the most pervasive challenge is the oracle problem~\cite{oracle-problem}. State-of-the-art tools often treat HTTP status codes as failure oracles, but this is unreliable. For instance, a \error response indicates an exception, which may be caused by many factors rather than a violation of specified requirements.

To address these challenges, we introduce \icepick, a framework for black-box testing of RESTful APIs that uses \tla~\cite{original-tla}, a formal language for specifying and designing complex software systems~\cite{tla}, together with its model checker, TLC~\cite{tlc, microsoft-tla}. We leverage TLC's State Space Graph (SSG) to systematically generate test suites covering modelled behaviour. To address the oracle problem, we propose \glacier, a contract specification language for RESTful APIs based on first-order logic. \glacier enables developers to enrich API specifications with practical properties for test generation, transforming specifications into valuable testing artefacts. Beyond providing useful test oracles, \glacier assigns precise semantics to the API and adapts to any extensible RESTful API specification language.

Recognising that writing formal specifications imposes additional development overhead, we provide a tool that automatically generates \glacier contracts from OAS files using HTTP semantics and RESTful conventions. Additionally, we offer a catalogue~\cite{glacier-catalogue} of standard formulae definitions for common properties such as referential integrity constraints.

As noted, while we illustrate the technique using microservice architectures and RESTful APIs, it should also apply to systems with a well-defined programmatic interface.

Following an overview of our solution, we introduce key background concepts and survey related work on the automated testing of RESTful APIs and on model-checking-based test generation. We then present the paper's contributions:

\begin{itemize}
  \item \textbf{\glacier}~\cite{glacier-catalogue}: a contract specification language for RESTful APIs based on first-order logic and REST semantics, with an automated contract generator and a catalogue of reusable formulae for common \mbox{API properties.}
  \item \textbf{\icepick framework}: a model checking-based approach for generating comprehensive test suites from \tla specifications and \glacier contracts.
  \item \textbf{Empirical evaluation} using the EvoMaster Benchmark (EMB)~\cite{emb}. 
  \item \textbf{Replication package}~\cite{rep-pack} with all source-code and specifications. 
\end{itemize}

\section{Overview}
\label{sec:overview}

\begin{figure}
  \centering 
  \includegraphics[width=0.45\textwidth]{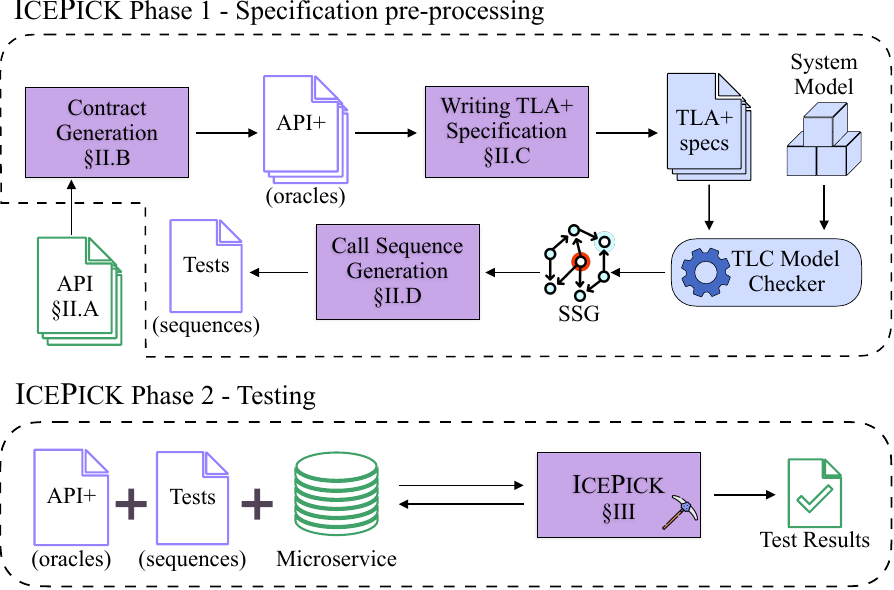}
  \caption{\icepick framework overview.}
  \label{fig:icepick-overview}
\end{figure}

This section overviews the \icepick framework using a tournament management system. The Tournaments Application comprises three APIs, \code{Tournaments}, \code{Players}, and \code{Enrolments}, enabling creation, update, and removal of these resources. \icepick's workflow has two phases: specification preprocessing and systematic testing, illustrated in Figure~\ref{fig:icepick-overview}.

In the first phase, we generate operation-specific contracts from API specifications and combine them into an extended specification to produce a \tla service abstraction. We then manually design the system model and check it with TLC to obtain the SSG. In the second phase, we traverse the SSG to generate call sequences, execute them against the system under test, and check observed behaviour against the specification using automatically generated test data~\cite{msc-thesis}. The process concludes with a detailed test report supporting reproducibility.

\subsection{OpenAPI Specification}
\label{subsec:oas}

The OAS describes RESTful APIs at the interface level: endpoints, input/output schemas, and status codes. While widely adopted, it neither models system state nor captures behavioural semantics, though it is extensible. In \icepick, we exploit this extensibility to attach operation-level contracts and invariants supporting test generation and oracle checking.

Listing~\ref{lst:post-player} shows the OAS description for \code{postPlayer} in the \code{Players} API, detailing its endpoint (line~\ref{lst:post-player-endpoint}), HTTP verb (\code{post}), a human-readable summary, unique identifier (line~\ref{lst:post-player-id}), request-body schema reference (line~\ref{lst:post-player-request-body}), and a successful response definition (lines~\ref{post-player-res-start}–\ref{post-player-res}). This information supports the generation of syntactically valid calls but is insufficient to determine whether the observed behaviour is semantically correct, motivating the contracts in~§\ref{subsec:contract-gen}.

\begin{lstlisting}[
  language=yaml,
  label={lst:post-player},
  caption={OAS description of \code{\footnotesize postPlayer} in the Tournament Application.},
  captionpos=b,
  escapechar=|
]
/players: |\label{lst:post-player-endpoint}|
  post: 
    summary: +Add a new player.+
    operationID: +postPlayer+ |\label{lst:post-player-id}|
    requestBody: |\label{lst:post-player-request-body}|
      content: 
        application/json: 
          schema: 
            $ref: +#/components/schemas/Player+
      required: +true+
    responses:  |\label{post-player-res-start}|
      200: 
        description: +Successful insertion.+
        content: 
		      application/json: 
            schema:
              $ref: +#/components/schemas/Player+ |\label{post-player-res}| 
\end{lstlisting}			      

\subsection{Contract Generation}
\label{subsec:contract-gen}
This module generates operation-level contracts (pre- and postconditions) from the OAS using CRUD (Create, Read, Update, Delete) semantics, expressed in a domain-specific language called \glacier. The generator produces an extended specification that retains the original OAS structure while incorporating contracts and invariants, which serve as executable oracles during testing.

This follows the established principle of using contracts as verification artefacts~\cite{contracts, heckel, dai, fm-book}, instantiated here for RESTful APIs via CRUD-inferred pre- and postconditions evaluated during black-box execution.

Listing~\ref{lst:post-player-contract} shows the inferred contract for \code{postPlayer}. 
The \code{requires} clause captures the expected precondition that the player does not yet exist (\notfound), while the \code{ensures} clause captures the expected postcondition that the player exists after insertion (\ok). The \code{@} symbol is a reference to ``this'' operation, enabling contracts to access the current request and response.

\begin{lstlisting}[
  language=yaml,
  label={lst:post-player-contract},
  caption={\code{\footnotesize postPlayer} operation contract.},
  captionpos=b,
  escapechar=|
]
/players: 
 POST: 
  operationID: +postPlayer+
  requires: 
  - +res_code(GET /players/req_body(@){pid}) = 404+
  ensures: 
  - +res_code(GET /players/req_body(@){pid}) = 200+
  - +req_body(@) = res_body(@)+|\label{post-player-extended}| 
\end{lstlisting}

Generated contracts are sound and automatically inferred, but intentionally partial: they may omit service-specific semantics. A manually added predicate can express application-specific consistency constraints (e.g., referential integrity) between tournament enrolments and the players registered in each tournament, which are not generally inferable from CRUD semantics alone. To support manual completion, we provide a catalogue of reusable \glacier predicates for common API properties~\cite{glacier-catalogue}. For instance, \code{postPlayer} can be strengthened by adding the predicate on line~\ref{post-player-extended} of Listing~\ref{lst:post-player-contract}, which validates that the returned payload matches the submitted input -- a semantic check derivable from the OAS definition that the operation receives and returns a player resource. 

In contrast, \code{deletePlayer} returns the player on success but does not include a player resource as input. Instead, it takes the player identifier as a path parameter. Consequently, the same equality predicate cannot be evaluated directly from the operation input. To address this, \glacier provides a \code{prev} operator that allows predicates to refer to values obtained before the operation executes. Listing~\ref{lst:delete-player-contract} shows the inferred contract for \code{deletePlayer}, using \code{prev} to compare the deleted resource with the value observed prior to deletion.

\begin{lstlisting}[
  language=yaml,
  label={lst:delete-player-contract},
  caption={\code{\footnotesize deletePlayer} operation contract.},
  captionpos=b,
  escapechar=|
]
/players/{pid}: 
 DELETE: 
  operationID: +deletePlayer+
  requires: 
  - +res_code(GET /players/{pid}) = 200+
  ensures: 
  - +res_code(GET /players/{pid}) = 404+
  - +req_body(@) = prev(res_body(GET /players/{pid}))+
\end{lstlisting}

Finally, \glacier supports quantification over collections, enabling expressive invariants. Listing~\ref{lst:inv} shows an invariant stating that, for every tournament, the number of enrolled players never exceeds the tournament capacity.

\begin{lstlisting}[
  language=yaml,
  label={lst:inv},
  caption={Invariant for the Tournaments API.},
  captionpos=b,
  escapechar=|
]
invariants: 
- +for t in res_body(GET /tournaments) :- 
	res_body(GET /tournaments/{t.tid}/players).len |$\leq$|
 res_body(GET /tournaments/{t.tid}/capacity) +
\end{lstlisting}

The previous examples introduce \glacier's main constructs, while Listing~\ref{lst:grammar} summarises its grammar. A formula is either a quantified statement over a collection or a boolean expression. Quantified formulas support both universal and existential quantification over multiple variables. Boolean expressions combine comparisons with standard logical operators, including implication (\code{=>}). The built-in functions \code{req\_body}, \code{res\_body}, and \code{res\_code} apply to the current operation (\code{@}) or an explicit HTTP request. The \code{prev} operator captures a value before execution, allowing postconditions to refer to the prior state. An optional suffix selects a field or applies a function such as \code{len}. The full grammar appears in~\cite{glacier-catalogue}.

\begin{lstlisting}[
  language=yaml,
  label={lst:grammar},
  caption={\glacier grammar.},
  captionpos=b, 
  basicstyle=\ttfamily\slshape\footnotesize\color{black},
  keywordstyle=\ttfamily\slshape\footnotesize\color{black},
  commentstyle=\ttfamily\slshape\footnotesize\color{black},
  stringstyle=\ttfamily\slshape\footnotesize\color{black},
  identifierstyle=\ttfamily\slshape\footnotesize\color{black},
  numberstyle=\ttfamily\slshape\footnotesize\color{black},
  literate={*}{{{\color{black}*}}}1,
  escapeinside={(*@}{@*)}
]
(*@\textcolor{purple-tla}{formula}@*) ::=  
  | ('for' | 'exists') vars (':-' | ':') formula
  | expr (boolOp expr)*
(*@\textcolor{purple-tla}{vars}@*) ::= varID 'in' call (',' vars)?
(*@\textcolor{purple-tla}{expr}@*) ::= 
  | ('res_code' | 'res_body' | 'req_body') '(' ('@' | method url) ')' suffix?
  | 'prev' '(' call ')'
  | expr comparator expr
(*@\textcolor{purple-tla}{suffix}@*) ::= '{' field '}' | '.' func
\end{lstlisting}

\subsection{\tla Specification}
\label{subsec:tla}
\tla provides a mathematical foundation for specifying systems~\cite{tla}, combining temporal logic and set theory. Paired with TLC~\cite{tlc, microsoft-tla}, it supports exhaustive state-space exploration and automated verification of safety and liveness properties. In \icepick, we use \tla to define a conservative abstraction that retains only state-relevant information for call sequence generation and oracle checking of a given system.

A \tla specification comprises three core elements: constants that define finite domains, variables that represent the abstract system state, and actions that describe state transitions~\cite{model-checking-guided}. In our example, each resource type is represented by a finite set of identifiers, and the system state is captured by maps from identifiers to simplified records. 
Operational contracts become action guards and next-state constraints, while global \glacier invariants become TLC-checked state predicates. We introduce an auxiliary boolean variable \code{final} marking terminal states (those in which all operations in a valid test sequence have been exercised) to guide SSG traversal in §\ref{subsec: algorithm}. The abstraction retains exactly the information needed to decide whether operations are enabled, how they change abstract state, and whether the invariants used by \icepick can be evaluated.

\subsubsection*{States and Types}
\label{subsubsec:tla-state} 

Listing~\ref{lst:tla-constants-variables} defines the constant domains (player, tournament, and enrolment identifiers, plus a finite set \code{N} of positive naturals for capacities) and the state variables (three resource maps and the terminal-state flag). Each constant is constrained by an \code{ASSUME} clause requiring non-empty identifier sets and a finite set \code{N} (not shown for brevity). Here, \code{EmptyMap} denotes the empty resource map, \code{IsMap} checks whether a variable is a map with the expected key and value domains, and \code{TS} identifies terminal states.

\begin{lstlisting}[
  language=tla,
  label={lst:tla-constants-variables},
  caption={\tla  constants and variables.},
  captionpos=b,
  escapechar=|
]
CONSTANTS PID, TID, EID, N
VARIABLES tournaments, players, enrolments, final
\end{lstlisting}  

Resources are represented as records omitting schema fields irrelevant to reachability, as shown in Listing~\ref{lst:tla-entities}. The initial state sets all resource maps to empty and \code{final} to \code{FALSE} (no lifecycles have completed yet).

\begin{lstlisting}[
  language=tla,
  label={lst:tla-entities},
  caption={\tla resource types and initial state.},
  captionpos=b
]
PlayerType     == [ts: SUBSET TID]
TournamentType == [ps: SUBSET PID, c: N]  
EnrolmentType  == [pid: PID, tid: TID]  

Init ==
  /\ players     = EmptyMap
  /\ tournaments = EmptyMap
  /\ enrolments  = EmptyMap
  /\ final = FALSE
\end{lstlisting}  

\subsubsection*{Invariants} 
Type correctness is enforced by a type invariant that verifies that resource maps are keyed by their identifiers and that \code{final} is a boolean. Beyond type correctness, we translate \glacier invariants into \tla predicates. Listing~\ref{lst:tla-typeinv} shows the type invariant checked by TLC: all state variables are maps except \code{final}, which is a boolean. 

\begin{lstlisting}[
  language=tla,
  label={lst:tla-typeinv},
  caption={Type invariant.},
  captionpos=b
]
TypeInv ==
  /\ IsMap(PID, PlayerType, players)
  /\ IsMap(TID, TournamentType, tournaments)
  /\ IsMap(EID, EnrolmentType, enrolments)
  /\ final \in BOOLEAN 
\end{lstlisting}  

Listing~\ref{lst:tla-inv1} specifies the invariant that the number of enrolled players never exceeds the tournament capacity, which model checking verifies for all reachable states.

\begin{lstlisting}[
  language=tla,
  label={lst:tla-inv1},
  caption={Tournaments capacity invariant.},
  captionpos=b
]
Inv1 == \A tid \in TID :
 tid \in DOMAIN tournaments => 
  |tournaments[tid].ps| <= tournaments[tid].c
\end{lstlisting} 

\subsubsection*{Operations as \tla actions}
Each state-mutating API operation becomes a \tla action, with pre- and postconditions derived from \glacier contracts. Listing~\ref{lst:tla-post-player} shows the complete action for \code{postPlayer}: the precondition requires that the player does not exist; the postcondition ensures it does after insertion. Line~\ref{line:tla-post-player-players} inserts the new player record, line~\ref{line:tla-post-player-f} updates the terminal-state flag, and line~\ref{line:tla-post-player-unchanged} specifies that tournaments and enrolments remain unchanged.

\begin{lstlisting}[
  language=tla,
  label={lst:tla-post-player},
  caption={\tla\ \code{\footnotesize postPlayer} operation definition.},
  captionpos=b, 
  escapechar=&
]
Requires(pid) == pid \notin DOMAIN players
Ensures(pid)  == pid \in DOMAIN players'

postPlayer(pid) == 
 /\ Requires(pid)
 /\ players' = MapPut(players, pid, [ts |-> {}])&\label{line:tla-post-player-players}&
 /\ Ensures(pid)
 /\ final' = TS(players', tournaments, enrolments)&\label{line:tla-post-player-f}&
 /\ UNCHANGED <<tournaments, enrolments>>&\label{line:tla-post-player-unchanged}&
\end{lstlisting}

\subsubsection*{Remarks}
Our \tla model is lifecycle-oriented and intentionally abstract, focusing on operations that induce state transitions (\code{post} and \code{delete}). Read-only operations are not considered, and update operations (\code{put} and \code{patch}) that modify abstracted fields are handled at the call sequence level~(§\ref{sec:puts}).

\subsection{Call Sequence Generation Algorithm}
\label{subsec: algorithm}
The SSG can be modelled as a directed graph $G=(V,E)$, where $V$ is the set of states explored by TLC and $E$ is the set of state transitions corresponding to API operations. It has a single initial state \code{i}, multiple final states, and a super-final sink \code{f} connecting all final states: for every final state $u$, $(u,\mbox{\code{f}})\in E$. Moreover, every state lies on some path from \code{i} to \code{f}, and the graph has no parallel edges.

We developed a custom coverage-guided algorithm based on breadth-first search (BFS)~\cite{cormen} to generate call sequences, described by paths from \code{i} to \code{f}. We chose the BFS-based approach since it produces the shortest paths from the initial state to the super-final state, thereby minimising the length of generated call sequences and reducing the number of API requests required during testing. The set of all selected call sequences achieves transition coverage, in the sense that every transition belongs to some selected path. The algorithm operates in three stages: first, it collects paths from \code{i},  classifying them as complete (reaching \code{f}) or incomplete; then, it collects a path from each state to \code{f}, used in the last step to complete incomplete paths. 

To simplify notation, states are denoted by integers from \code{i} to \code{f}. The digraph $G=(V,E)$ is implemented with two arrays of adjacency-lists:
$\mbox{\code{outAdj[$u$]}} = \{v \mid (u,v)\in E\}$ and
$\mbox{\code{inAdj[$v$]}} = \{u \mid (u,v)\in E\}$, where $u,v \in V$.

\subsubsection*{Stage 1: Collecting paths from the initial state} 
Algorithm~\ref{alg:paths_to} starts by creating and initialising some structures: the sets of complete and incomplete paths and the FIFO queue start empty, no state has been found, and there is no path from \code{i} to any state (lines~\ref{paths_to:init_1}--\ref{paths_to:init_2}). The traversal starts at \code{i}: the zero-length path from \code{i} to \code{i} is stored, \code{i} is enqueued and is marked as found (lines~\ref{paths_to:analyseI_1}--\ref{paths_to:analyseI_2}).
In the while loop, the algorithm dequeues a state and explores all of its outgoing transitions. 
For each transition, it extends a copy of the current path by appending the destination state
(line~\ref{paths_to:append}). Then, three distinct cases can hold. If the destination state is \code{f}, since \code{f} has been declared found in line~\ref{paths_to:init_3}, the new path is added to the set of complete paths. Otherwise, either the destination state has not yet been found, in which case the new path will be extended further, the state is enqueued and marked as found, or the destination state has already been enqueued, and the new path is added to the set of incomplete paths. When the queue is empty, the sets of complete and incomplete paths are returned.

\begin{algorithm}
\caption{Collects paths from \code{i} (complete and incomplete).}
\label{alg:paths_to}
\begin{algorithmic}[1]
  \Statex \textbf{input:}
    \Statex \hspace{0.2em} \code{i} : initial state, \quad \code{f} : final state
    \Statex \hspace{0.2em} \code{outAdj} : 
                           states adjacent to each state by outgoing transitions
  \Statex \textbf{output:} 
    \Statex \hspace{0.2em} \code{cmp} : set of complete paths
    \Statex \hspace{0.2em} \code{inc} : set of incomplete paths

  \Function{PathsTo}{}
    \State \code{cmp} $\gets \varnothing$, \quad 
           \code{inc} $\gets \varnothing$ 
           \label{paths_to:init_1}
    \State \code{paths[i$\dots$f]} $\gets \bot$
    \State \code{fifo} $\gets \varnothing$, \quad    
           \code{found[i$\dots$f]} $\gets \bot$ 
           \label{paths_to:init_2}
    \State \code{found[f]} $\gets \top$
           \label{paths_to:init_3}
    \State \code{paths[i]} $\gets$ \code{(i)}
           \label{paths_to:analyseI_1}
    \State \Call{Enqueue}{\code{fifo}, \code{i}}, \quad
           \code{found[i]} $\gets \top$ 
           \label{paths_to:analyseI_2}
    \While{\code{fifo} $\neq \varnothing$}
      \State \code{u} $\gets$ \Call{Dequeue}{\code{fifo}}
      \ForAll{\code{v} $\in$ \code{outAdj[u]}}
        \State \code{pathV} $\gets$ 
               \Call{Append}{\Call{Clone}{\code{paths[u]}}, \code{(v)}}
               \label{paths_to:append}
        \If{\code{found[v]} $= \bot$}
          \State \code{path[v]} $\gets$ \code{pathV}
          \State \Call{Enqueue}{\code{fifo}, \code{v}}, \quad
                 \code{found[v]} $\gets \top$  \label{paths_to:found}
        \ElsIf{\code{v} $=$ \code{f}} 
          \State \code{cmp} $\gets$ \code{cmp} $\cup$ \code{\{pathV\}}
        \Else
          \State \code{inc} $\gets$ \code{inc} $\cup$ \code{\{pathV\}}
        \EndIf
      \EndFor
    \EndWhile
    \State \Return \code{cmp,\ inc}
\EndFunction
\end{algorithmic}
\end{algorithm}

Let \code{cmp} and \code{inc} be the returned sets. Since every state lies on some path from \code{i} to \code{f} and each transition is analysed once, being integrated in some path, set \code{cmp} $\cup$ \code{inc} covers all transitions. Besides, the size of \code{cmp} $\cup$ \code{inc} is, at most, the number of transitions, as no transition is shared across paths. 

\subsubsection*{Stage 2: Collecting paths to the super-final state} 
Algorithm~\ref{alg:paths_from} collects, for each state, a shortest path from that state to \code{f}. Like Algorithm~\ref{alg:paths_to}, it also executes a breadth-first search, but now the traversal starts at \code{f}, transitions are analysed from the destination to the origin state, and new paths are built by prefixing the origin to a copy of the current path (line~\ref{paths_from:append}).

\begin{algorithm}
\caption{Collects a path from each state to \code{f}.}
\label{alg:paths_from}
\begin{algorithmic}[1]
  \Statex \textbf{input:}
    \Statex \hspace{0.2em} \code{i} : initial state, \quad \code{f} : final state
    \Statex \hspace{0.2em} \code{inAdj} : 
                           states adjacent to each state by incoming transitions 
  \Statex \textbf{output:} 
    \Statex \hspace{0.2em} \code{paths} : a path from each state to \code{f}

  \Function{PathsFrom}{}
    \State \code{paths[i$\dots$f]} $\gets \bot$
    \State \code{fifo} $\gets \varnothing$, \quad    
           \code{found[i$\dots$f]} $\gets \bot$
    \State \code{paths[f]} $\gets$ \code{(f)}
    \State \Call{Enqueue}{\code{fifo}, \code{f}}, \quad
           \code{found[f]} $\gets \top$ 
    \While{\code{fifo} $\neq \varnothing$}
      \State \code{v} $\gets$ \Call{Dequeue}{\code{fifo}}
      \ForAll{\code{u} $\in$ \code{inAdj[v]}}
        \If{\code{found[u]} $= \bot$}
          \State \code{paths[u]} $\gets$ 
                 \Call{Append}{\code{(u)}, \Call{Clone}{\code{paths[v]}}}
                 \label{paths_from:append}
          \State \Call{Enqueue}{\code{fifo}, \code{u}}, \quad
                 \code{found[u]} $\gets \top$
        \EndIf
      \EndFor
    \EndWhile
    \State \Return \code{paths}
\EndFunction
\end{algorithmic}
\end{algorithm}

\subsubsection*{Stage 3: Selecting call sequences for diversity}
Since the set $\mbox{\code{cmp}} \cup \mbox{\code{inc}}$ already covers all transitions, the main goal of Algorithm~\ref{alg:select_sequences} is to extend the incomplete paths. The set of \code{selected} paths starts with \code{cmp}, which contains only complete paths (line~\ref{select_sequences:init}).
For each incomplete path collected in the first phase, $p=\mbox{\code{i}} \, u_1 \, u_2 \cdots u_{n} \, v$ (for some $n\geq 1$), a new complete path is created by first deleting the last state from  $p$ and then appending the path from $v$ to \code{f}, as computed in the second phase (lines~\ref{select_sequences:delete}--\ref{select_sequences:append}). The new path is added to the selected paths.

\begin{algorithm}
\caption{Selects the call sequences from the SSG.}
\label{alg:select_sequences}
\begin{algorithmic}[1]
  \Function{SelectSequences}{\code{ssg} : state-space graph}
    \State (\code{cmp,\ inc}) $\gets$ 
           \Call{PathsTo}{\code{ssg.i}, \code{ssg.f}, \code{ssg.outAdj}}
    \State \code{from} $\gets$ 
           \Call{PathsFrom}{\code{ssg.i}, \code{ssg.f}, \code{ssg.inAdj}}
    \State \code{selected} $\gets$ \code{cmp}
           \label{select_sequences:init}
    \ForAll{\code{p} $\in$ \code{inc}} \label{line:for_incomplete}
      \State \code{v} $\gets$ \Call{DeleteLast}{\code{p}} 
             \label{select_sequences:delete}
      \State \code{q} $\gets$ \Call{Append}{\code{p}, \Call{Clone}{\code{from[v]}}}
             \label{select_sequences:append}
      \State \code{selected} $\gets$ \code{selected} $\cup$ \code{\{q\}}
    \EndFor
  \State \Return \code{selected}
  \EndFunction
\end{algorithmic}
\end{algorithm}

\subsubsection*{Dealing with updates}
\label{sec:puts}
As discussed in §\ref{subsec:tla}, our \tla abstraction models only \code{post} and \code{delete} operations. We therefore insert \code{put} calls into the generated sequences by randomly augmenting the output of Algorithm~\ref{alg:select_sequences} with zero to three consecutive \code{put} operations per created resource. This emulates common usage patterns (e.g., repeated profile edits), supports idempotency checks (identical consecutive updates with the same input), and exercises overwrite behaviour under successive updates with different inputs.

To keep sequences valid, updates must occur after creation and before deletion. Accordingly, we identify matching \code{post}/\code{delete} pairs and insert \code{put} operations between them. When a created resource has no corresponding \code{delete} in a sequence, we insert \code{put} operations at arbitrary positions, but always after the creation call.

\subsubsection*{Remarks}
If there are parallel transitions, the algorithm may not achieve 100\% transition coverage because each path is represented as a sequence of states, thereby losing information about which operation caused the transition. This representation choice is sufficient for full state coverage, but may lose transition identity when multiple operations connect the same abstract states; we quantify this effect in §~\ref{sec:ecal-call-seq-gen}. The assumption that there were no parallel transitions comes from the fact that in RESTful APIs, different operations typically have distinct preconditions or effects, so operations that appear to transition between the same abstract states usually differ in the data they manipulate (e.g., different resource IDs). The algorithm's output is a collection of call sequences, i.e., invocations of the API operations, each consisting of an operation identifier and its parameters. Since these sequences are generated by TLC's model checking, they contain model values rather than concrete API parameters.

\section{\icepick}
\label{sec:icepick}
The \icepick tool comprises five core components that enable systematic black-box testing of RESTful APIs. The \code{input\; generator} produces random data conforming to the specification schemas. The \code{state\; emulator} maintains an abstract representation of the microservice state observed during testing, tracking which resources exist and their key properties. The \code{evaluator} checks whether \glacier predicates hold at each execution point by interpreting contract clauses against observed responses and emulated state. The HTTP \code{request\; handler} manages all communication between the tester and the microservice, including request construction, transmission, and response capture. Finally, the \code{tester} orchestrates these components within the overall testing algorithm, coordinating the execution of call sequences and aggregating results.

\subsection{Testing Algorithm}

Algorithm~\ref{alg:test_all} depicts \icepick's testing procedure. It iterates over all generated call sequences, treating each as an independent test scenario. For each sequence, the algorithm resets the state emulator (line~\ref{emul_reset}) to enforce isolation between runs and prevent state carryover from influencing subsequent sequences. It then executes each operation in sequence, aggregating the outcomes into the overall result. 

\begin{algorithm}
\caption{\icepick's testing algorithm.}
\label{alg:test_all}
\begin{algorithmic}[1]
  \Function{Test}{\code{seqs}: call sequences, \code{spec}: extended spec}
    \State \code{results} $\gets \varnothing$
    \ForAll{\code{seq} $\in$ \code{seqs}}
      \State \code{emulator}.\Call{Reset}{}       \label{emul_reset}
      \ForAll {\code{op} $\in$ \code{seq}}
        \State \code{res} $\gets$ \Call{TestOperation}{\code{op}, \code{spec}, \Call{IsLast}{\code{op}}}
        \State \code{results} $\gets$ \code{results} ++ \code{res}
      \EndFor
    \EndFor
  \State \Return \code{results}
  \EndFunction
\end{algorithmic}
\end{algorithm}

Algorithm~\ref{alg:test_single} presents \icepick's algorithm for testing a single operation. It starts by verifying that system invariants hold in the pre-execution state (line~\ref{alg:invs}). This check ensures the system maintains global consistency properties before attempting the operation.

\begin{algorithm}
\caption{\icepick's operation testing algorithm.}
\label{alg:test_single}
\begin{algorithmic}[1]
  \Statex \textbf{input:}
    \Statex \hspace{0.2em} \code{op}: operation to test
    \Statex \hspace{0.2em} \code{spec}: \glacier specification 
    \Statex \hspace{0.2em} \code{last}: indicates if \code{op} is the last one in the call sequence
  \Statex \textbf{output:}
    \Statex \hspace{0.2em} \code{result}: test outcome

  \Function {TestOperation}{}
    \State \code{verb } $\gets$ \code{op.verb}, \quad\quad\quad\quad\code{schema} $\gets$ \code{op.schema}
    \State \code{reqs } $\gets$ \code{op.reqs}, \quad\quad\quad\quad\code{ens} $\gets$ \code{op.ens}
    \State \code{tlaID} $\gets$ \code{op.parameters}, \ \code{data} $\gets$ $\emptyset$
    \State \code{invs } $\gets$ \code{evaluator}.\Call{EvalInv}{\code{spec.invs}} \label{alg:invs}
    
    \Switch {\code{verb}}
      \Case {\code{POST}}
        \State \code{data} $\gets$ \code{generator}.\Call{Generate} {\code{schema}} \label{alg:post_gen}
        \State \code{pre } $\gets$ \code{evaluator}.\Call{EvalPre}{\code{reqs}, \code{data}}
        \State \code{resp} $\gets$ \code{handler}.\Call{Request}{\code{verb}, \code{data}}
        \If {\code{resp} $=$ \code{OK}}
          \State \code{emulator}.\Call{Add}{\code{tlaID}, \code{data}} \label{post:emul}
        \EndIf
      \EndCase
      \Case {\code{DELETE}}
        \State \code{data} $\gets$ \code{emulator}.\Call{Recycle}{\code{tlaID}} \label{del:recycle}
        \If {data $= \varnothing$} \Return \code{NOT\_TESTED}\EndIf \label{del:not_tested}
        \State \code{pre } $\gets$  \code{evaluator}.\Call{EvalPre}{\code{reqs}, \code{data}}
        \State \code{resp} $\gets$ \code{handler}.\Call{Request}{\code{verb}, \code{data.id}}
        \If {\code{resp} $=$ \code{OK}}
          \State \code{emulator}.\Call{Delete}{\code{tlaID}, \code{data}} \label{del:emul}
        \EndIf
      \EndCase
      \Case {\code{PUT}}
        \State \code{data} $\gets$ \code{emulator}.\Call{Recycle}{\code{tlaID}} \label{put:recycle}
        \If {data $= \varnothing$} \Return \code{NOT\_TESTED}\EndIf \label{put:not_tested}
        \State \code{data} $\gets$ \code{generator}.\Call{Generate}{\code{schema}}
        \State \code{pre } $\gets$ \code{evaluator}.\Call{EvalPre}{\code{reqs}, \code{data}}
        \State \code{resp} $\gets$ \code{handler}.\Call{Request}{\code{verb}, \code{data.id}}
        \If {\code{resp} $=$ \code{OK}}
          \State \code{emulator}.\Call{Update}{\code{tlaID}, \code{data}} \label{put:emul}
        \EndIf
      \EndCase
    \EndSwitch
    \If{\code{resp} $=$ \code{OK}}
      \State \code{pos} $\gets$ \code{evaluator}.\Call{EvalPos} {\code{ens}, \code{data}, \code{resp}}
    \EndIf
    \If{\code{last}}
      \State \code{invs} $\gets$ \code{evaluator}.\Call{EvalInv}{\code{spec.invs}}
    \EndIf
    \State \Return \Call{Result}{\code{resp}, \code{invs}, \code{pre}, \code{pos}}
  \EndFunction
\end{algorithmic}
\end{algorithm}

The algorithm then branches on the operation's HTTP verb, handling data generation and recycling differently in each case:

\noindent\textbf{POST operations} create new resources, so the input generator produces fresh data conforming to the operation's request body schema (line~\ref{alg:post_gen}). After evaluating preconditions using the generated data, the algorithm issues the HTTP request. If the service responds with a success code (\success), the created resource is added to the state emulator (line~\ref{post:emul}), making it available for subsequent operations that may reference it.

\noindent\textbf{DELETE operations} remove existing resources, so the algorithm attempts to recycle previously generated data (line~\ref{del:recycle}) from the state emulator, identified here by \code{tlaID}. If no matching data exists, typically because the corresponding \code{post} failed, the operation is marked as \code{NOT\_TESTED} (line~\ref{del:not_tested}) and skipped. Otherwise, preconditions are evaluated, the \code{delete} request is issued with the resource identifier, and upon success, the resource is removed from the state emulator (line~\ref{del:emul}).

\noindent\textbf{PUT operations} update existing resources, requiring both recycling of the target resource identifier (line~\ref{put:recycle}) and generation of new data for the update payload. Like \code{delete}, if the required resource is unavailable in the state emulator, the operation is marked as \code{NOT\_TESTED} (line~\ref{put:not_tested}). Otherwise, the algorithm generates new data conforming to the schema, evaluates the preconditions, issues the \code{put} request, and updates the emulated state upon success (line~\ref{put:emul}).

After executing the operation, if the service does not return a server error (as indicated by a \error response), the algorithm evaluates the postconditions using the operation's ensures clauses.
If the operation is the last in the sequence, the algorithm performs a final invariant check to ensure that system-wide properties still hold upon completion.

\noindent\textbf{GET operations} retrieve resources and are pure, i.e., they have no side-effects on the service's state. These operations allow clients to inspect the system state and serve a dual purpose in \icepick. First, they enable evaluation of \glacier contracts by providing observable state information for preconditions, postconditions, and invariants. Second, because \code{get} operations do not trigger state transitions in our \tla model, we do not test them as standalone operations. Instead, we use their responses as instrumentation and as the ground truth for validating the behaviour of state-mutating operations. 

\subsection{Result Classification}

\icepick classifies each operation test into one of four categories: \code{OK}, \code{WARN}, \code{ERR}, or \code{NOT\_TESTED}. This classification scheme provides feedback beyond simple pass-or-fail verdicts, reflecting the complexity of contract-based testing, where violations may be ambiguous.

\code{NOT\_TESTED} indicates the state emulator lacks the required data to proceed with the request. This typically occurs when a \code{delete} or \code{put} operation references a resource that a \code{post} operation failed to create earlier in the sequence. For all other cases, Table~\ref{tab:op_outcomes} shows how the framework classifies operations based on three factors: whether preconditions held before execution, whether postconditions and invariants held after execution, and the HTTP response code returned by the service.

\begin{table}[h]
  \centering
  \setlength{\tabcolsep}{3pt}
  \renewcommand{\arraystretch}{0.8}
  \caption{Operation Results.}
  \label{tab:op_outcomes}
  \begin{tabular}{c c c c c}
    \toprule
    \textbf{Preconditions} & \textbf{Postconditions} & \textbf{Invariants} & \textbf{Response Code} & \textbf{Result}\\
    \midrule 
    \multirow{4}{*}{T} & \multirow{2}{*}{T} & T & \success{} / $\neg$\success & \code{OK} / \code{ERR} \\
                       &                     & F & -- & \code{ERR} \\
    \cmidrule{2-5}
                       & \multirow{2}{*}{F} & T & -- & \code{ERR} \\
                       &                     & F & \clienterr{} / \success & \code{WARN} / \code{ERR} \\
    \midrule
    \multirow{4}{*}{F} & \multirow{2}{*}{T} & T & -- & \code{WARN} \\
                       &                     & F & -- & \code{ERR} \\
    \cmidrule{2-5}
                       & \multirow{2}{*}{F} & T & \clienterr{} / $\neg$\clienterr & \code{OK} / \code{ERR} \\
                       &                     & F & \clienterr{} / \success & \code{OK} / \code{ERR} \\
    \bottomrule
  \end{tabular}
\end{table}

An operation test is classified as \code{OK} when the observed behaviour is consistent with the contract, i.e., preconditions, postconditions, and invariants hold, and the response code matches the outcome defined by the specification. The result of a test is \code{ERR} when there is a clear contract violation or an inconsistent response code. This includes several fault patterns. For instance, when preconditions hold, but postconditions or invariants do not, or when the response code indicates success while the contract evaluation indicates a failure. Finally, \code{WARN} handles ambiguous cases where the outcome is suspicious but not definitively wrong. Specifically, when preconditions hold, postconditions and invariants do not hold, and the service returned a \clienterr status code. This is suspicious because the service rejected what appeared to be valid input per the specification, yet the contract violation suggests the system entered an unexpected state. In this same scenario, if the service replies with a \success response code, we are definitely dealing with an error, since the service claims success while violating the contract. In the latter warning scenario, the contract is violated, so by definition~\cite{design-by-contract} we have no guarantee of the service's state when the operation executes, and we conservatively classify the result as a warning.

This classification scheme enables \icepick to distinguish among failure types, allowing more effective debugging. The framework generates detailed test reports that identify the violated contracts, the sequence of operations leading to the failure, and the specific preconditions, postconditions, or invariants that were not met, providing developers with valuable diagnostic information.

\section{Evaluation}
Our experiments aim to evaluate both the framework and the proposed testing algorithm. Accordingly, this section is divided into three parts: SSG generation with TLC, call sequence generation, and testing, each with its own research questions. 

All experiments were conducted on four identical Debian GNU/Linux 12 machines, each equipped with two AMD EPYC 7343 processors, for a total of 32 physical cores and 64 hardware threads, 128 GiB of DDR4 3200 MHz main memory, and a 450 GB SSD as the primary storage device. The machines were connected via a 2$\times$10 Gbps network.

Table~\ref{table:suts} describes the systems used to evaluate our framework. All services except the Tournaments System are drawn from the EMB~\cite{emb}. Petstore~\cite{petstore} is the canonical example API for OAS; E-commerce service is a shopping cart application built as a technology-learning project~\cite{new-emb}; and Features-Service manages configurable products and their features. For our experiments, we upgraded their OAS to version 3.1.0 and then validated and Dockerised each application for deployment.

\begin{table}
  \centering
  \setlength{\tabcolsep}{4pt}
  \renewcommand{\arraystretch}{1}
  \caption{Systems under test description.}
  \label{table:suts}
  \begin{tabular}{l * {4}{r}}
    \toprule
    \textbf{System} & \textbf{Endpoints} & \textbf{Operations} & \textbf{Parameters} & \textbf{Schemas}\\
    \midrule 
    Tournaments       & 11 & 19 & 15 & 4  \\ 
    Swagger-Petstore  & 13 & 19 & 17 & 6  \\
    E-Commerce        & 22 & 28 & 33 & 25 \\
    Features-Service  & 10 & 16 & 33 & 8  \\
    \bottomrule
  \end{tabular}
\end{table}

\subsection{Applicability Analysis -- Requirements for effective use} 

\icepick is designed for testing systems that adhere to REST architectural principles and HTTP specifications. The approach works best when services properly implement standard HTTP semantics, including accurate status codes, resource-oriented design, and stateless interactions. These properties enable \icepick to effectively infer system state and detect invariant violations. During our evaluation, we found that many existing benchmarks used in prior work deviate significantly from these principles, making them unsuitable for demonstrating \icepick's strengths. 
This observation highlights an important consideration for the research community: as REST testing techniques become more sophisticated, we need evaluation benchmarks that reflect proper REST architectural patterns. The EMB~\cite{emb} represents a significant step forward, providing a valuable curated collection of systems for evaluation. However, our experience with two systems from this benchmark -- Features-Service and E-Commerce API -- shows that creating benchmarks comprehensively reflecting REST architectural principles is difficult, and even carefully curated collections may include systems with varying degrees of compliance. 

When generating the extended specification files, we found several issues. Our generator (§\ref{subsec:contract-gen}) invokes the official OAS validator, which revealed that several specifications were invalid. This allowed us to identify and correct issues such as missing required parameters, duplicated parameters, invalid URLs, undefined path parameters, invalid or unused schema definitions, missing response codes, and absent tag descriptions. While manually resolving most of these issues was straightforward, assigning meaningful tags required understanding the semantics of each operation and the resources it manipulates. We analysed the specification files and organised operations into groups (i.e., tags) based on their domain responsibilities and the resources they manage. 

For Features-Service, we added tags \mbox{\code{Products}, \code{Features}}, \code{Configurations}, and \code{Constraints} aligned with the domain model and to separate concerns by resource.
We also found signs that the API description is generated from source code: the file included an invalid schema name, \code{Map<string,\ Link>}, which we removed as an unused reference.
We then inferred a \tla specification for this service from the OAS-generated contracts. However, testing revealed that all \code{post} operations lacked request-body definitions, which may be common practice, but does not follow the standard. RFC 9110\cite{rfc-post} (HTTP Semantics) specifies that a \code{post} is used for creating new resources that have yet to be identified by the origin server. Without request body schemas, clients cannot know what data is expected. This was compounded by Features-Service always returning success status codes \success, even for malformed or unprocessable requests. This violates basic HTTP principles -- status codes should reflect the actual outcome of a request. With missing request-body definitions and success responses for everything, there is no way to distinguish valid operations from invalid ones, or to assess whether the tool's generated requests are actually correct. 

For the E-Commerce API, our generator could not infer contracts because the API violates basic HTTP and REST conventions. For example, \code{delete} requests include a body (discouraged by RFC~9110) and their URIs do not conform to common design guidelines, which prevents us from deriving a meaningful \tla specification; therefore, we excluded them. Instead, we evaluate on Tournaments, Swagger-Petstore, and Features-Service, which better support the assessment of our approach.

Regarding the manual effort of writing \tla specifications, we can say both Tournaments and Petstore specifications are comparable in size. While we did not measure authoring time, the specifications appear to be small enough for a \tla practitioner to manage manually, though this remains to be validated empirically. Since reducing manual specification effort is a requirement for broader applicability, automating \tla generation from enriched OAS files is a primary direction for future work, as mentioned in §~\ref{sec:concl}.

Overall, our applicability analysis indicates that \icepick is most effective for APIs that expose resource-oriented operations, provide accurate request/response schemas, and are compliant with HTTP status-code semantics.

\subsection{SSG Generation}
Next, we investigate the scalability of SSG generation using TLC and the relationship between model size and the resulting state space. We aim to determine whether TLC can generate SSGs for realistic API models within acceptable time and storage constraints. If model checking becomes impractical for model sizes beyond the trivial, the entire approach fails. As such, we address the following research question:

\begin{description}
  \item[\textbf{RQ1}] What is the scalability limit on model size for which graph generation performance remains acceptable?
\end{description}

\begin{figure}
  \centering 
  \includegraphics[width=0.492\textwidth]{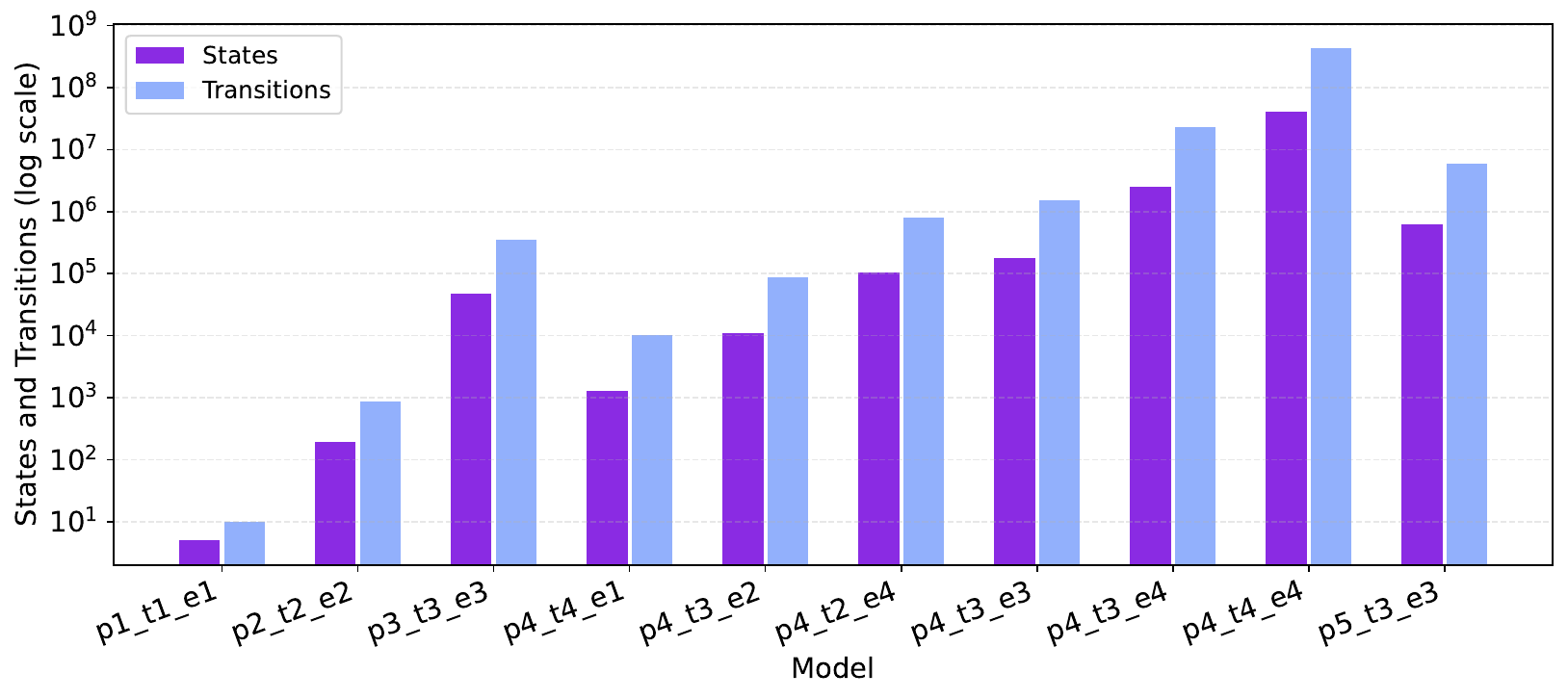}
  \caption{Model size vs. number of states and transitions generated by TLC for the Tournaments API.}
  \label{fig:model-size}
\end{figure}

We ran TLC for the Tournaments System with configurations with total model sizes ranging from 3 to 12, e.g., 4 players, 4 tournaments, and 4 enrolments (\code{p4\_t4\_e4}). Figure~\ref{fig:model-size} compares the model size with the number of states/transitions generated by the model checker. As expected, larger models yield more states, though models of the same size can differ by several orders of magnitude (e.g., \code{p3\_t3\_e3} vs. \code{p4\_t4\_e1}). This is expected: fewer enrolments reduce the number of possible states by restricting the \code{player}/\code{tournament} combinations for each enrolment. We also observe that for relatively small models, e.g., \code{p4\_t3\_e4}, the SSG has more than 2 million states and 23 million transitions.

With state spaces reaching millions of states even for small models, one might consider leveraging TLC's state-reduction techniques, such as symmetric model value sets. However, this approach does not apply here: in our examples, each model value maps directly to a concrete API resource, making the values intrinsically different rather than interchangeable.

Analysing TLC-generated SSG files, we observed substantial duplication in nodes and edges that would degrade call sequence generation performance, so we added a preprocessing step to remove redundancy. Table~\ref{table:dot_clean} reports size reduction.

Thanks to our pre-processing stage, the file-size reductions range from 30\% to 99.4\% across all 17 configurations; in over 75\% of the cases, more than 50\% of the original size is eliminated, and in three of the four Swagger-Petstore configurations, the reduction exceeds 90\%. Pre-processing significantly reduces the size of the SSG representation, lowering storage costs and enabling faster subsequent traversals.

\begin{table}
  \centering
  \caption{SSG generation -- execution time and size reduction.}
  \label{table:dot_clean}
  \begin{tabular}{V c T r r r}
    \toprule
    & \textbf{Model} & \textbf{Exec. TLC} & \textbf{Original} & \textbf{Clean} & \textbf{Reduction} \\
    \midrule
    \multirow{10}{*}{\rotatebox[origin=c]{90}{\textsc{Tournaments}}} 
    & \code{p1\_t1\_e1} &   $<$ 01s & 3 K   &   2 K & 33.3 \% \\
    & \code{p2\_t2\_e2} &   $<$ 01s & 357 K & 166 K & 53.5 \% \\
    & \code{p3\_t3\_e3} & 00min 21s & 243 M &  62 M & 74.5 \% \\
    & \code{p4\_t4\_e1} &   $<$ 01s & 1.6 M & 569 K & 64.6 \% \\
    & \code{p4\_t3\_e2} & 00min 05s & 61 M  &  15 M & 75.6 \% \\
    & \code{p4\_t2\_e4} & 00min 45s & 497 M & 143 M & 71.8 \% \\
    & \code{p4\_t3\_e3} & 01min 49s & 1.3 G & 262 M & 79.2 \% \\
    & \code{p4\_t3\_e4} & 36min 09s & 21 G  &   4 G & 81.0 \% \\
    & \code{p4\_t4\_e4} & 16h 16min & 467 G &  72 G & 84.6 \% \\
    & \code{p5\_t3\_e3} & 07min 48s & 5.5 G & 974 M & 82.7 \% \\
    \midrule
    \multirow{4}{*}{\rotatebox[origin=c]{90}{\textsc{Petstore}}}
    & \code{u1\_p1\_o1} &   $<$ 01s &   5 K & 3,5 K & 30.0 \% \\
    & \code{u2\_p2\_o2} &   $<$ 01s & 794 K &  70 K & 91.2 \% \\
    & \code{u3\_p3\_o3} & 00min 05s & 112 M & 2,3 M & 97.9 \% \\
    & \code{u4\_p4\_o4} & 11min 25s &  15 G &  95 M & 99.4 \% \\
    \midrule
    \multirow{3}{*}{\rotatebox[origin=c]{90}{\textsc{Feat.}}}
    & \code{p1\_c1\_f1} &   $<$ 01s & 3.7 K & 3.7 K &    0 \% \\
    & \code{p2\_c2\_f2} &   $<$ 01s & 1.3 M & 710 K & 45.4 \% \\
    & \code{p3\_c3\_f3} & 06min 20s & 5.2 G & 1.9 G & 63.6 \% \\
    \bottomrule
  \end{tabular}
\end{table}

Experiments show that SSG generation scales to moderate model sizes: for the Tournaments API, configurations up to size $~$11 ($\leq$21 GB) finish in under 40min, but size 12 balloons to 467 GB and 16h~16min; for Petstore, the largest tried (15 GB) finishes in 11min~25s. On our hardware, DOT/SSG outputs of roughly 5-21 GB can be produced within an hour, after which build time becomes impractical for quick testing.

\begin{callout}
  \begin{description}
  \item[\textbf{RQ1}] TLC scales to moderate model sizes -- configurations within $\sim$5-21 GB complete within an hour. Beyond this (e.g., 467 GB for Tournaments with 12 model values), generation becomes impractical (16+ hours).
  \end{description} 
\end{callout}

\subsection{Call Sequence Generation}
\label{sec:ecal-call-seq-gen}

This section evaluates the scalability and effectiveness of our Java implementation of the call-sequence generation algorithm for SSG exploration. Even if TLC successfully generates an SSG, the call sequence generation algorithm must be able to process it within available memory; understanding these limits prevents users from attempting infeasible configurations. We configured the JVM with 100 GB heap and ran the algorithm for each model configuration, addressing the following research questions:

\begin{description}
  \item[\textbf{RQ2}] How large can the SSG grow while our algorithm still runs without exhausting our memory limits?
  \item[\textbf{RQ3}] What configurations should we test, considering our time and memory constraints? 
\end{description}

We consider RQ1, RQ2, and RQ3 to establish \icepick's practical feasibility by determining: 
\begin{enumerate*}[label=(\roman*)]
  \item whether TLC can generate SSGs within acceptable time/storage limits;
  \item whether the call sequence algorithm can process these graphs within memory constraints; and 
  \item which model configurations are tractable for real-world testing.
\end{enumerate*}
Together, they bridge the gap between theoretical soundness and practical applicability.

Table~\ref{table:call_seq_gen} summarises the results of our experiments per model, reporting the resulting SSG's size (in terms of number of states and transitions), and the number of distinct paths generated. We also show the graph coverage obtained from the generated path set, in terms of both visited states and visited transitions (full results in the replication package~\cite{rep-pack}). We only execute tests for one- and two-value-per-resource configurations, since three values yield an infeasible number of call sequences (tens of thousands), which translates into millions of API requests.

For all SSGs, the algorithm quickly achieves full state coverage, effectively enumerating all possible paths. We were not able to generate call sequences for the \code{p3\_c3\_f3} of Features-Service (FS.) on our setup due to memory constraints. Petstore (Pets.) configurations achieve only 74\% to 84\% transition coverage, unlike the 100\% obtained for Tournaments (Tour.) and Features-Service. This is due to the parallel-transition assumption discussed in §\ref{subsec: algorithm}: some Petstore operations share the same abstract states but differ in data manipulation (e.g., \code{createUser} and \code{createUserWithList} both transition from a state where the user does not exist to one where it does). Since paths are represented only as state sequences, the algorithm selects one transition per state pair and misses the other.

Regarding SSG size limits, the largest configuration we could run within our memory constraints was \code{p3\_t3\_e3} for the Tournaments system, with 9 model values, 46K states, and 349K transitions.

\begin{callout}
  \begin{description}
    \item[\textbf{RQ2}] With a 100 GB heap, the algorithm handles SSGs with up to $\sim$46K states and $\sim$349K transitions.
    \item[\textbf{RQ3}] Only one- and two-value-per-resource configurations are practical; three yield tens of thousands of sequences (millions of API requests), making testing infeasible.
  \end{description}
\end{callout}

\begin{table}
  \centering
  \setlength{\tabcolsep}{0.4em}
  \renewcommand{\arraystretch}{1}
  \caption{Call sequence generation.}
  \label{table:call_seq_gen}
  \begin{tabular}{V c r r r r r r r r r}
    \toprule
      &  
      & \multicolumn{2}{c}{\textbf{Graph Size}}
      & \multicolumn{3}{c}{\textbf{Path Size}}
      & \multicolumn{2}{c}{\textbf{Graph Cov.}}
      \\
    \cmidrule(lr){3-4} \cmidrule(lr){5-7} \cmidrule(lr){8-9} 
      & \textbf{Model} 

      & \textbf{States}      
      & \textbf{Trans.}     

      & \textbf{Min}
      & \textbf{Max}
      & \textbf{Avg}  

      & \textbf{States}     
      & \textbf{Trans.}

      & \textbf{Paths}
      \\
      \midrule
      \multirow{3}{*}{\rotatebox[origin=c]{90}{\textsc{Tour.}}}
      & \code{p1\_t1\_e1} & 6    & 10   & 6 & 11 & 9  & 100\% & 100\% & 7       \\
      & \code{p2\_t2\_e2} & 193  & 872  & 9 & 27 & 17 & 100\% & 100\% & 721     \\
      & \code{p3\_t3\_e3} & 46K  & 349K & 9 & 39 & 24 & 100\% & 100\% & 312K    \\

      \midrule
      \multirow{3}{*}{\rotatebox[origin=c]{90}{\textsc{Pets.}}}
      & \code{u1\_p1\_o1} & 7    & 19    & 3 & 6  & 4 & 100\%  & 84.2\% & 15    \\
      & \code{u2\_p2\_o2} & 73   & 574   & 5 & 10 & 7 & 100\%  & 81.2\% & 507   \\
      & \code{u3\_p3\_o3} & 1361 & 19.9K & 7 & 13 & 9 & 100\%  & 74.4\% & 18.5K \\
      
      \midrule
      \multirow{2}{*}{\rotatebox[origin=c]{90}{\textsc{FS.}}}
      & \code{p1\_c1\_f1} & 12   & 36    & 2 & 7  & 5  & 100\% & 100\% & 32    \\
      & \code{p2\_c2\_f2} & 4490 & 49.8K & 6 & 16 & 10 & 100\% & 100\% & 46K   \\
      
    \bottomrule
  \end{tabular}
\end{table}

\subsection{Testing}

In this section, we investigate \icepick's ability to uncover faults through systematic state-space exploration, and assess whether enriching automatically generated contracts with manually specified semantic properties enhances fault detection beyond basic HTTP semantics.
As such, we aim to answer the following research questions: 

\begin{description}
  \item[\textbf{RQ4}] How effective is \icepick in finding faults? 
  \item[\textbf{RQ5}] Does the manual extension of specifications result in an increased number of detected faults?
\end{description}

To evaluate fault-detection, we injected three faults into the Tournaments system: 
\begin{enumerate*}[label=(\roman*)]
  \item \code{deletePlayer}  does not remove the player;
  \item \code{deleteTournament} removes a random tournament; and
  \item \code{deleteEnrolment} fails to update the Tournament's players collection, violating referential integrity.
\end{enumerate*}
The latter was discovered during preliminary testing rather than intentionally injected.

\begin{table}
  \centering
  \setlength{\tabcolsep}{0.7em}
  \renewcommand{\arraystretch}{1}
  \caption{Test Results.}
  \label{table:test-results}
  \begin{tabular}{V l r r r c}
    \toprule
      & \textbf{Model} 
      & \textbf{Exec. Time}
      & \code{WARN}
      & \code{ERR}
      & \code{NOT\_TESTED}
      \\
    \midrule
    \multirow{3}{*}{\rotatebox[origin=c]{90}{\textsc{Tour.}}}
      & \code{p1\_t1\_e1}           & 00min 02s & 0  & 0 & 0 \\
      & \code{p2\_t2\_e2}           & 02min 20s & 0  & 0 & 0 \\
      & \code{p1\_t1\_e1\_ext}      & 00min 14s & 0  & 0 & 0 \\
    \midrule
    \multirow{3}{*}{\rotatebox[origin=c]{90}{\textsc{T. ERR}}}
      & \code{p1\_t1\_e1\_ERR}       & 00min 02s & 8    & 5    & 0  \\
      & \code{p2\_t2\_e2\_ERR}       & 02min 44s & 1658 & 459  & 49 \\
      & \code{p1\_t1\_e1\_ERR\_ext}  & 00min 17s & 7    & 5    & 0  \\

      \midrule
      \multirow{2}{*}{\rotatebox[origin=c]{90}{\textsc{\,Pets.}}}
      & \code{u1\_p1\_o1} & 00min 01s & 0 & 0 & 3   \\
      & \code{u2\_p2\_o2} & 00min 11s & 0 & 0 & 72  \\
      
      \midrule
      \multirow{1}{*}{\rotatebox[origin=c]{90}{\textsc{FS.}}}
      & \code{p1\_c1\_f1} & $<$ 01s   & 0 & 97 & 50  \\
      
    \bottomrule
  \end{tabular}
\end{table}

Table~\ref{table:test-results} shows the test execution results for the Tournaments, Petstore, Features-Service, and the faulty Tournaments system, across different model configurations. Columns \code{WARN} and \code{ERR} count the number of individual operation executions that were classified as errors or warnings, according to \icepick's result classification scheme in Table~\ref{tab:op_outcomes}. Each operation tested during the execution of all call sequences contributes one result, which is categorised as \code{OK}, \code{WARN}, \code{ERR}, or \code{NOT\_TESTED}. So, e.g., the 97 errors of Features-Service imply 97 operation executions violated their contracts or exhibit inconsistent behaviour. 

All Tournaments configurations completed successfully (under 3 minutes). For Petstore, both completed without warnings or errors, but left 3 and 72 operations untested: instances of \code{deleteUser} in sequences where users were created via \code{createUserWithList} rather than \code{postUser}. Since \code{createUserWithList} stores users as an array while \code{deleteUser} expects a single object, a distinction not inferable from the specification, \icepick marks these as \code{NOT\_TESTED}.

Features-Service was tested with a single configuration due to its non-compliance with REST architectural principles and HTTP semantics. \icepick found 97 errors and 50 untested operations in the smallest possible configuration for this service, which completed in under one second. All operations marked as errors are \code{post}, and all operations marked as \code{NOT\_TESTED} are operations depending on resource creation (e.g., \icepick cannot test the deletion of a product if the product could not be added to the system). These operations are classified as errors because they lack definitions in their request bodies. 

\icepick detected all injected faults in the Tournaments system (T. ERR), even in the smallest configuration. However, some faults were semantically subtle. A representative example is the referential-integrity fault in \code{deleteEnrolment}. \icepick first creates a state in which a player is enrolled in a tournament, then executes the deletion and checks the resulting state against the specification. Although the faulty implementation returns request bodies lacking information and produces the same HTTP response as the correct one, it leaves the tournament's player collection unchanged, resulting in an inconsistent state where the enrolment is removed, but the tournament still references the player. This shows why stateful, model-guided exploration is necessary: the fault only becomes visible after establishing the relevant multi-resource state.

The extended specification version shows comparable detection capability to the generated version for the small model, suggesting that even automatically generated contracts are effective at catching the injected defects. However, the extended specification produced one fewer warning than the generated one. Log analysis revealed a \code{postEnrolment} operation failed both its pre and postconditions in the extended specification test, resulting in an \code{OK} classification when the service returned a \clienterr response (according to Table~\ref{tab:op_outcomes}). The postcondition failure was due to tournament capacity: random data generation created a tournament with a small capacity that was exhausted by previous enrolments. In contrast, the baseline test randomly generated a tournament with a larger capacity, so the postcondition passed, resulting in a \code{WARN} classification for the same test.

\icepick proved effective at finding faults. It detected all three injected faults in the Tournaments system, even when using the smallest model configuration. It also uncovered 97 errors in Features-Service, all related to missing request body definitions and violations of HTTP semantics. However, \icepick's effectiveness depends on how well the API adheres to REST principles: well-designed APIs, such as Petstore and Tournaments, enable comprehensive testing, whereas poorly compliant systems, such as Features-Service, immediately reveal structural problems that block deeper exploration. In essence, \icepick was effective in two different ways: it detected injected behavioural faults in the REST-compliant Tournaments system, and it exposed severe specification/HTTP-semantics deficiencies in Features-Service that prevented meaningful deeper exploration.

Manually extending specifications did not substantially improve fault detection over automatically generated contracts in our injected-fault setting: both detected all Tournaments faults, and the only observed difference was a warning caused by random data generation. This indicates that CRUD- and HTTP-based inferred contracts were sufficient for the fault types represented in our injections. However, this should not be interpreted as evidence that manual contracts are unnecessary: manually added \glacier predicates are expected to be more beneficial for domain-specific semantic faults that are not captured by generic CRUD behaviour and were not exercised by our injected faults.

\begin{callout}
\begin{description}
  \item[\textbf{RQ4}] \icepick detected all injected Tournaments faults and 97 errors in Features-Service, though its effectiveness depends on REST compliance.
  \item[\textbf{RQ5}] In our injected-fault setting, manual extensions did not improve detection; inferred contracts sufficed for the exercised fault patterns.
\end{description}
\end{callout}

\section{Threats to Validity}

\noindent\textbf{Internal Validity.} 
The correctness and completeness of the generated \glacier contracts pose a threat: if a predicate is incorrect, or a contract is overly permissive or restrictive, the observed effectiveness may reflect specification quality rather than \icepick's fault-finding capability -- i.e., overconstrained specifications may lead to missing the exploration of valid call sequences, while underconstrained ones may lead to the generation of invalid call sequences~\cite{politano}. 
To mitigate this we manually reviewed all generated contracts, checking them against the OAS textual descriptions and CRUD operation semantics. Similarly, \tla specifications may diverge from the real system due to abstraction, so we sanity-checked all cases and iteratively refined them whenever observed API behaviour differed from SSG-implied behaviour. Moreover, our injected faults primarily target behaviours that can be expressed by inferred CRUD- and HTTP-level contracts, potentially understating the added value of manually specified domain-specific predicates. A further internal validity threat is the framework's implementation: defects in any component could bias results. While we tested all components, we cannot guarantee the absence of faults. To mitigate this and support reproducibility, we publicly release the framework implementation and the case studies in our replication package~\cite{rep-pack}.

\par\noindent\textbf{External Validity.} 
We evaluated \icepick on only three systems due to the scarcity of high-quality open-source APIs.
To mitigate this, we plan to evaluate \icepick with more EMB~\cite{emb} systems. Finally, the OAS files we used may be cleaner and more complete than real-world specifications. In practice, specifications can be outdated -- i.e., the specification file does not evolve at the same pace as the implementation -- or inconsistent, which can affect the framework's effectiveness. 

\section{Related Work}
\label{sec:related_work}

\tla has seen significant industrial adoption beyond Amazon~\cite{amazon-tla}, including use at MongoDB for verifying distributed reconfiguration protocols~\cite{mongodb-tla}. While these applications focus on verifying distributed system designs, our work adapts \tla to black-box API testing, a challenging problem requiring operational dependencies inference, generation of schema-compliant inputs, and the definition of test oracles beyond HTTP status codes. While existing methods address these issues to varying degrees, key gaps remain; this work addresses them by combining contract-based verification with advanced call sequence generation. 
 
Several tools infer operation dependencies to generate valid call sequences. \rtg~\cite{restestgen} builds a static operation dependency graph (ODG) from inter-operation data flows; MoREST~\cite{morest} extends this with a dynamic, runtime-updated resource property graph (RPG) to enable longer sequences, but requires exploratory calls; RESTler~\cite{restler} combines fuzzing with CRUD semantics and feedback from prior requests to incrementally infer dependencies. JepREST\cite{jeprest} targets concurrency correctness by verifying linearisability of REST operations under concurrent execution, complementing \icepick's focus on sequential state-space coverage.

Beyond operation dependencies, some tools focus on parameter dependencies. RESTest~\cite{restest} analyses inter-parameter relationships to construct valid calls and produce both nominal and faulty tests. It uses richer oracles beyond \error status codes by also checking OAS conformance and whether valid inputs receive successful responses. RestCT~\cite{restct} applies combinatorial testing across operations and parameters, generating operation sequences with constrained sequence covering arrays and input assignments with classical covering arrays. It uses CRUD semantics and URI hierarchy to handle inter-operation constraints. RestCT provides systematic t-way coverage of operation combinations, i.e., every combination of $t$ operations (or $t$ parameter values) appears at least once across the generated test sequences, which fundamentally differs from our state-space coverage over the formal behavioural model.

EvoMaster~\cite{evomaster1, evomaster2} treats test generation as a search-based optimisation problem, using the Many Independent Objective (MIO) algorithm to evolve tests that maximise coverage and fault detection. It features two execution modes: the white-box mode~\cite{evomaster1} instruments the SUT collecting coverage metrics to guide the search; in the black-box mode~\cite{evomaster2}, it relies on CRUD semantics and URI hierarchy to structure test sequences.

ModelFuzz~\cite{modelfuzz} uses \tla to guide fuzzing of distributed systems implementations. Instead of exploring program input spaces as traditional fuzzers do, it explores event schedules in distributed systems, using abstract model state coverage as feedback to guide test generation. This approach maps implementation events to model actions and uses a controlled model checker to track which abstract states have been covered, and to mutate schedules that reach new states. While ModelFuzz is designed for distributed system testing rather than RESTful APIs, it shares the underlying philosophy of leveraging higher-level abstractions to define test oracles and to guide exploration beyond output validation.

Existing tools often misclassify \error responses as defects, generate mostly random inputs, and lack robust mechanisms to verify behaviour beyond status codes. These approaches either improve dependency inference, coverage, or input generation, but still rely on limited oracles and do not support systematic state-space exploration. The literature also includes additional approaches, such as specification-based techniques~\cite{spec-based}, metamorphic testing strategies~\cite{metamorphic}, and the reinforcement learning-based exploration proposed by ARAT-RL~\cite{arat-rl}. Although these methods address related aspects of API testing, they target different problem settings and rely on testing paradigms that differ substantially from our approach.

\section{Conclusion and Future Work}
\label{sec:concl}

This paper presents \icepick, a framework for systematic black-box testing of RESTful APIs using model checking to achieve comprehensive state-space coverage. \icepick combines \tla specifications with the TLC model checker to generate test suites that provably cover the modelled behavioural space. It derives call sequences through a coverage-guided BFS traversal guaranteeing full state coverage and, when the abstraction has no parallel transitions, full transition coverage.
Our evaluation highlights both the potential and the boundaries of the approach. For REST-compliant APIs, \icepick uncovers subtle multi-operation faults (e.g., referential integrity violations) that tools relying solely on HTTP status code oracles would miss. However, REST compliance is not just a desirable property but a hard requirement: as our experience with Features-Service and E-Commerce APIs shows, when APIs deviate from REST principles, \icepick may be unable to infer contracts, making testing infeasible before it begins. When testing is feasible, the framework produces reproducible test reports -- an advantage over search-based or fuzzing approaches whose outputs vary across runs. At the same time, our results clarify where \icepick's current limits lie. The approach requires APIs to adhere to REST principles, manual effort to write \tla specifications, and careful model sizing to stay within tractable time and memory bounds. These are not fundamental barriers, but they do define the conditions under which \icepick is most effective.

\noindent \textbf{Future Work.} Several directions are worth exploring. First, our evaluation only covers three systems, so testing EMB systems would strengthen the external validity of our findings. Second, writing \tla specifications currently requires manual effort and formal methods expertise; automating this step to derive the full \tla specification directly from the extended file with \glacier contracts is a primary direction for future work. Third, adapting \icepick to handle partially REST-compliant APIs, by relaxing assumptions about status code semantics or resource-oriented URI design, would broaden its applicability to real-world systems. Finally, when SSG contains parallel transitions, our algorithm cannot guarantee full transition coverage; extending the call-sequence generation to track transition identity, rather than representing paths purely as state sequences, would close this gap while preserving strong coverage guarantees.

Overall, \icepick shows that combining model checking with automated contract inference and executable test oracles can generate comprehensive, reproducible API test suites with strong coverage guarantees.

\section*{Acknowledgments}
The authors were funded by NOVA LINCS (UID/\allowbreak04516/\allowbreak2025), FCT I.P., and the EU Horizon TaRDIS project (Grant Agreement No.~\allowbreak101093006).

\balance
\bibliographystyle{IEEEtran}
\bibliography{bibliography}

\end{document}